\documentclass{pasa}%

\usepackage{graphicx}
\usepackage{aas_macros}
\usepackage{hyperref}
\usepackage{color}

\newcommand{\bwt}{\begin{widetext}}
\newcommand{\ewt}{\end{widetext}}
\newcommand{\beq}{\begin{equation}}
\newcommand{\eeq}{\end{equation}}
\newcommand{\bea}{\begin{eqnarray}}
\newcommand{\eea}{\end{eqnarray}}

\begin{document}

\title[The Equation of State for the Nucleonic and Hyperonic Core of Neutron Stars]%
{The Equation of State for the Nucleonic and Hyperonic Core of Neutron Stars}
\author[Tolos et al.]{Laura Tolos$^{1,2}$, Mario Centelles$^3$ and Angels Ramos$^3$
\affil{$^1$Institute of Space Sciences (CSIC-IEEC), Campus Universitat Aut\`onoma de Barcelona, Carrer de Can Magrans, s/n, 08193 Cerdanyola del Vall\`es,
Spain}
\affil{$^2$Frankfurt Institute for Advanced Studies, Goethe University Frankfurt, Ruth-Moufang-Str. 1,
60438 Frankfurt am Main, Germany} 
\affil{$^3$Departament de F\'{\i}sica Qu\`antica i Astrof\'{\i}sica and Institut de Ci\`encies del Cosmos (ICCUB), Universitat de Barcelona,
Mart\'{\i} i Franqu\`es 1, 08028 Barcelona, Spain}
}

\jid{PASA}
\doi{10.1017/pas.\the\year.xxx}
\jyear{\the\year}

\hypersetup{colorlinks,citecolor=blue,linkcolor=blue,urlcolor=blue}

\begin{frontmatter}
\maketitle

\begin{abstract}
We reexamine the equation of state for the nucleonic and hyperonic inner core of 
neutron stars that satisfies the 2$M_{\odot}$ observations as well as the recent 
determinations of stellar radii below 13 km, while fulfilling the saturation properties of nuclear matter and finite nuclei  together with the constraints on the high-density nuclear pressure
coming from heavy-ion collisions. 
The recent nucleonic FSU2R and hyperonic FSU2H models are updated in order to improve the behavior of pure neutron matter at subsaturation densities. The corresponding nuclear matter properties at saturation, the symmetry energy and its slope
turn out to be compatible with recent experimental and theoretical determinations.
We obtain the mass, radius and composition of neutron stars for the two updated models and study the impact on these properties of the uncertainties in the hyperon-nucleon couplings estimated from hypernuclear data. We find that the onset of appearance of each hyperon strongly depends on the hyperon-nuclear uncertainties, whereas the maximum masses for neutron stars differ by at most 0.1 $M_{\odot}$, although a larger deviation should be expected tied to the lack of knowledge of the hyperon potentials at the high densities present in the center of $2 M_\odot$ stars.
For easier use, we provide tables with the results from the FSU2R and FSU2H models for the equation of state and the neutron star mass-radius~relation.
\end{abstract}

\end{frontmatter}

\section{Introduction}
\label{sec:intro}

The equation of state (EoS) of matter inside neutron stars has received a lot of attention over the last decades \citep{Lattimer:2004pg,Lattimer:2006xb,Oertel:2016bki}. Besides black holes, neutron stars (usually observed as pulsars) are the most compact known objects in the universe. Their bulk features, such as mass and radius, strongly depend on the properties of matter in their interior and, hence, on the EoS.

With regards to mass determinations, the most precise measurements of masses are clustered around the Hulse-Taylor pulsar of 1.4$M_{\odot}$ \citep{Hulse:1974eb}. However,  accurate values of approximately 2$M_{\odot}$ have been determined very recently. This is the case of the PSR J1614-2230 of $M=1.97 \pm 0.04 M_{\odot}$ \citep{Demorest:2010bx} and the PSR J0348+0432 of $M=2.01\pm 0.04 M_{\odot}$ \citep{Antoniadis:2013pzd}. 

As for radii, precise determinations do not yet exist due to the difficulties in modeling the X-ray spectra emitted by the atmosphere of a neutron star \citep{Verbiest:2008gy,Ozel:2010fw,Suleimanov:2010th,Lattimer:2012xj,Steiner:2012xt, 
Bogdanov:2012md,Guver:2013xa,Guillot:2013wu,Lattimer:2013hma,Poutanen:2014xqa, 
Heinke:2014xaa,Guillot:2014lla,Ozel:2015fia,Ozel:2015gia,Ozel:2016oaf,
Lattimer:2015nhk}. Nevertheless, most of these analysis seem to favor small radii below 13 km \citep{Fortin:2014mya}. High-precision X-ray space missions, such as the on-going NICER (Neutron star Interior Composition ExploreR) \citep{2014SPIE.9144E..20A}, will shed some light by offering simultaneous measurements of masses and radii  \citep{Watts:2016uzu}, whereas neutron-star radii are expected to be measured with a precision 
of 1 km by means of gravitational-wave signals coming from neutron-star mergers  \citep{Bauswein:2011tp,Lackey:2014fwa}.

In view of these findings and future observations, it is opportune to analyze whether theoretical models for the EoS of dense matter can satisfy both the $2 M_\odot$ maximum mass constraint and radii below 13 km. Moreover, these models should fulfill the saturation properties \footnote{Saturation properties refer to the physical characteristics of infinite symmetric nuclear matter at the density $\rho_0$, where the energy per particle, $E/A$, presents a minimum.} of nuclear matter and finite nuclei (or atomic nuclei). On the one hand, in order to obtain small neutron star radii, a softening of the pressure of neutron matter, and hence of the nuclear symmetry energy, around 1-2 times saturation density $n_0$ ($n_0 \approx 0.16$~${\rm fm}^{-3}$) is required  \citep{Lattimer:2006xb,Tsang:2012se,Ozel:2016oaf}. On the other hand, the total pressure should be stiff enough in order to sustain 2$M_{\odot}$ neutron stars. Very few models can reconcile simultaneously both constraints (small radius and large masses) and, at the same time, produce a precise description of finite nuclei \citep{Jiang:2015bea,Horowitz:2000xj,Horowitz:2001ya,Chen:2015zpa,Sharma:2015bna}.

Furthermore, as density increases inside neutron stars, the transition from nuclear to hyperonic matter would be favoured energetically \citep{1960AZh....37..193A}. Thus, the EoS softens as new degrees of freedom, hyperons, appear \citep{Glendenning:1982nc} leading to smaller neutron stars masses, below the 2$M_{\odot}$ observations. This is known as the ``hyperon puzzle'', whose solution requires a new mechanism to stiffen the EoS: stiffer hyperon-nucleon and/or hyperon-hyperon interactions, repulsive three-body forces with hyperons, new hadronic degrees of freedom that push the onset of appearance of hyperons to higher densities or the phase transition to quark matter below the hyperon onset (see Ref.~\citep{Chatterjee:2015pua} and references herein).

In a recent paper  \citep{Tolos:2016hhl} we have obtained the EoS for the nucleonic and hyperonic inner core of neutron stars by reconciling the $2 M_{\odot}$ mass observations with the recent analyses of radii below 13 km for neutron stars. Moreover, we have fulfilled the saturation properties of nuclear matter and finite nuclei  \citep{Tsang:2012se,Chen:2014sca}  as well as  the recent constraints extracted from nuclear collective flow \citep{Danielewicz:2002pu} and kaon production \citep{Fuchs:2000kp,Lynch:2009vc} in heavy-ion collisions (HICs).  
The study was performed in the relativistic mean-field (RMF) theory for describing both the nucleon and hyperon interactions and the EoS of the neutron star core.
Two models were formulated, denoted as FSU2R (with nucleons) and FSU2H (with nucleons and hyperons), based on the nucleonic FSU2 model of \citep{Chen:2014sca}.  

In the present paper, we update the parameters of our two models in order to improve the behaviour of the EoS of pure neutron matter (PNM) at subsaturation densities by avoiding possible instabilities in the low-density region. We determine the properties at saturation of the modified interactions and we compare our results for the symmetry energy and the slope of the symmetry energy to recent experimental and theoretical determinations, while providing predictions for the neutron skin thickness of the $^{208}$Pb and $^{48}$Ca nuclei. Finally, we reinvestigate the mass-radius relationships for the two models, and estimate the impact on the neutron star masses, radii and composition of the uncertainties in the hyperon-nucleon couplings.

The paper is organized as follows. In Sec.~\ref{sec:formalism} we present the RMF model for the 
determination of the EoS in beta-equilibrated matter. In Sec.~\ref{sec:eos} 
we show the newly calibrated nucleonic FSU2R and hyperonic FSU2H models. Then, in 
Sec.~\ref{sec:stellar} we display the results for the mass-radius relationship for neutron stars and in Sec.~\ref{sec:hyperon} we estimate the impact on the stellar properties of the uncertainties in the hyperon-nucleon couplings. We 
finally summarize our results in Sec.~\ref{sec:summary}. Tables with numerical data of the EoSs are provided in the Appendix.


\section{Theoretical Framework}
\label{sec:formalism}

In the covariant field theory of hadronic matter, the baryons are treated as Dirac 
particles that interact through the exchange of mesons \citep{Serot:1984ey}. The 
formalism has been in wide use over the last four decades for describing the properties 
of the nuclear EoS and of finite nuclei in a relativistic quantum framework.
A contemporary formulation of the Lagrangian density of the theory 
\citep{Serot:1984ey,Serot:1997xg,Glendenning:2000,Chen:2014sca} may be 
written in terms of the contributions from the baryons ($b$), leptons ($l$=$e$, $\mu$), 
and mesons ($m = \sigma$, $\omega$, $\rho$, and $\phi$) as
\bea
{\cal L} &=& \sum_{b}{\cal L}_{b} + {\cal L}_{m}+ \sum_{l}{\cal L}_{l} , \nonumber \\
{\cal L}_{b}&=&\bar{\Psi}_{b}(i\gamma_{\mu}\partial^{\mu}-q_{b}\gamma_{\mu}A^{\mu}- 
m_{b}  \nonumber \\
&+&  g_{\sigma b}\sigma
-g_{\omega b}\gamma_{\mu}\omega^{\mu}-g_{\phi b}\gamma_{\mu}\phi^{\mu}-g_{\rho b}\gamma_{\mu}\vec{I}_b \, \vec{\rho \, }^{\mu} 
)\Psi_{b} , \nonumber \\[2mm] 
{\cal L}_{l}&=& \bar{\psi}_{l}\left(i\gamma_{\mu}\partial^{\mu}-q_{l}\gamma_{\mu}A^{\mu}
-m_{l}\right )\psi_{l} ,\nonumber \\[2mm] 
{\cal L}_{m}&=&\frac{1}{2}\partial_{\mu}\sigma \partial^{\mu}\sigma
-\frac{1}{2}m^{2}_{\sigma}\sigma^{2} - \frac{\kappa}{3!} (g_{\sigma N}\sigma)^3 - \frac{\lambda}{4!} (g_{\sigma N}\sigma)^4 \nonumber \\
&-& \frac{1}{4}\Omega^{\mu \nu} \Omega_{\mu \nu} +\frac{1}{2}m^{2}_{\omega}\omega_{\mu}\omega^{\mu}  + \frac{\zeta}{4!}   (g_{\omega N}\omega_{\mu} \omega^{\mu})^4 \nonumber \\
&-&\frac{1}{4}  \vec{R}^{\mu \nu}\vec{R}_{\mu \nu}+\frac{1}{2}m^{2}_{\rho}\vec{\rho}_{\mu}\vec{\rho \, }^{\mu} + \Lambda_{\omega} g_{\rho N}^2 \vec{\rho}_{\mu}\vec{\rho \,}^{\mu} g_{\omega N}^2 \omega_{\mu} \omega^{\mu} \nonumber \\
&-&\frac{1}{4}  P^{\mu \nu}P_{\mu \nu}+\frac{1}{2}m^{2}_{\phi}\phi_{\mu}\phi^{\mu} -\frac{1}{4} F^{\mu \nu}F_{\mu \nu} ,
\label{lagran}
\eea
where $\Psi_{b}$ and $\psi_{l} $ stand for the baryonic and leptonic Dirac fields, respectively.  The mesonic and electromagnetic field strength tensors are $\Omega_{\mu \nu}=\partial_{\mu}\omega_{\nu}-\partial_{\nu}\omega_{\mu}$, $\vec{R}_{\mu 
\nu}=\partial_{\mu}\vec{\rho}_{\nu}-\partial_{\nu}\vec{\rho}_{\mu} $, $P_{\mu \nu}=\partial_{\mu}\phi_{\nu}-\partial_{\nu}\phi_{\mu}$ and  $F_{\mu \nu}=\partial_{\mu}A_{\nu}-\partial_{\nu}A_{\mu}$. 
The isospin operator is represented by the vector $\vec{I}_b$.
The strong interaction coupling of a meson to a certain baryon is denoted by $g$ (with $N$ indicating nucleon)  
and the electromagnetic couplings by $q$, while the masses of the baryons, mesons, and leptons are denoted by~$m$.  

The coupling constants of the above Lagrangian encode in an approximate way the 
complicated nuclear many-body dynamics. 
The $g_{\sigma N}$ and $g_{\omega N}$ couplings of the isoscalar $\sigma$ 
and $\omega$ mesons to the nucleon determine the energy per 
particle and density of the nuclear matter saturation point, and, thus, are 
instrumental for the ground-state properties of finite nuclei. The $g_{\rho N}$ 
coupling of the isovector $\rho$ meson to the nucleon is key for the 
nuclear symmetry energy. Essentially, the symmetry energy measures the energy cost 
involved in changing all the protons into neutrons in nuclear matter 
\citep{Li:2014oda}. Therefore, the $g_{\rho N}$ coupling impacts on the properties of 
heavy neutron-rich nuclei and of neutron stars. The Lagrangian density 
(\ref{lagran}), moreover, incorporates self-interactions of the meson fields. The 
$\sigma$-meson self-interactions, with the $\kappa$ and $\lambda$ couplings, were 
introduced by \citep{Boguta:1977xi} and allowed for the first quantitatively successful 
descriptions of nuclear matter and finite nuclei within the relativistic theory.  
These couplings soften the EoS at moderate densities and allow one to obtain 
a realistic compressibility of nuclear matter \citep{Boguta:1977xi,Boguta:1981px} in 
agreement with the values extracted from experiments on nuclear giant resonances and heavy ion 
collisions.\footnote{Note that it has been suggested that the nuclear compressibility 
could be also inferred from gravitational wave observations of pulsar glitch recoveries \citep{Bennett:2010tm}} 
The quartic self-coupling $\zeta$ of the vector $\omega$ meson was introduced by
\citep{Bodmer:1991hz}. The $\zeta$ coupling must be nonnegative to prevent abnormal 
solutions of the vector field equation of motion \citep{Bodmer:1991hz,Mueller:1996pm}. 
It then implies an attractive nonlinear interaction that softens the EoS for high 
densities \citep{Bodmer:1991hz}, thereby directly affecting the structure and maximum 
mass of neutron stars \citep{Mueller:1996pm}.
Finally, a mixed interaction between the $\omega$ and $\rho$ mesons, with the coupling 
$\Lambda_{\omega}$, modulates the density dependence of the nuclear symmetry energy---which 
is related to the pressure of neutron matter---and influences the neutron radius 
of heavy nuclei and the radii of neutron stars \citep{Horowitz:2000xj,Horowitz:2001ya}.

The Dirac equations for the different baryons and leptons are obtained from the 
Lagrangian density (\ref{lagran}) as
\bea
&&(i\gamma_{\mu}\,\partial^{\mu}-q_{b}\,\gamma_{0}\,A^{0}-m^{*}_{b} \nonumber \\
&&-g_{\omega b}\, \gamma_{0} \, \omega^{0} -g_{\phi b} \,\gamma_{0}\, \phi^{0} 
-g_{\rho b}\, I_{3 b}\, \gamma_{0} \,\rho_3^{0}) \Psi_{b}=0 , \nonumber \\ 
&&\left(i\gamma_{\mu}\,\partial^{\mu}-q_{l}\,\gamma_{0} \,A^{0}-m_{l} \right) \psi_{l}=0 , \label{MFlep}
\eea
where the quantities
\beq
m^{*}_{b}=m_{b}-g_{\sigma b}\sigma \label{effmass} 
\eeq
denote the effective masses of the baryons.
Let us mention that only the time-like component of the vector fields and the 
third component of isospin have been written in Eq.~(\ref{MFlep}) due to the assumption 
of rotational invariance and charge conservation. The field equations of motion of the 
mesons follow from the respective Euler--Lagrange equations, see for example \citep{Serot:1984ey}.
Altogether, the theory leads to a set of coupled nonlinear field equations that involve 
strong couplings. The exact solution of these equations is extremely complicated if one 
attempts to quantize both the baryon fields and the meson fields. 
Physically, the baryons are the constituents of the nuclear medium, whereas the mesons 
are the carriers of the interaction between baryons. Thus, in order to be able to solve 
the equations of the theory, it is meaningful to replace the meson field operators by 
their expectation values, which then act as classical fields in which the baryons move. 
This approach is known as the relativistic mean-field theory \citep{Serot:1984ey}. 
Denoting the meson mean fields in uniform matter as 
$\bar \sigma= \langle \sigma \rangle$, $\bar\omega=\langle\omega^0\rangle$, $\bar\rho=\langle\rho_3^0\rangle$,
and $\bar \phi=\langle\phi^0\rangle$, 
the mesonic equations of motion in the mean-field approximation for the uniform medium are
\begin{eqnarray}
&&m_\sigma^2 \, \bar \sigma + \frac{\kappa}{2} g_{\sigma N}^3 \bar \sigma^2 + \frac{\lambda}{3!}  g_{\sigma N}^4 \bar \sigma^3 = \sum_{b} g_{\sigma b} n_b^s , \nonumber \\ 
&& m_\omega^2 \, \bar \omega + \frac{\zeta}{3!}  g_{\omega N}^4 \bar \omega^3 + 2 \Lambda_{\omega} g_{\rho N\,}^2  g_{\omega N}^2  \bar \rho^2 \bar \omega = \sum_{b} g_{\omega b} n_b , \nonumber \\ 
&& m_\rho^2 \,  \bar \rho + 2 \Lambda_{\omega} g_{\rho N}^2  g_{\omega N}^2  \bar \omega^2 \bar \rho= \sum_{b} g_{\rho b} I_{3 b} n_b , \nonumber \\ 
&& m_\phi^2 \bar \phi \, = \sum_{b} g_{\phi b} n_b ~, \label{eqphi}
\end{eqnarray}
where $I_{3 b}$ is the third component of the isospin of a given baryon, 
and we use the convention that for protons $I_{3 p}=+1/2$. 
The quantities 
\bea
n_b^s &=& \langle \bar \Psi_b \Psi_b \rangle , \nonumber \\
n_b   &=& \langle \bar \Psi_b\gamma^0 \Psi_b\rangle ,
\eea
are, respectively, the scalar and vector densities for the $b$ baryon. 
In terms of the baryonic and leptonic Fermi momenta, $k_{Fb}$ and $k_{Fl}$, and of the 
respective Fermi energies 
\bea
E_{Fb}&=&\sqrt{k_{Fb}^{2}+m^{*2}_{b}} \  , \nonumber \\
E_{Fl}&=& \sqrt{k_{Fl}^{2}+m^{2}_{l} } \ , 
\eea
the scalar and vector densities for the baryons and the vector densities for the leptons are expressed as 
\bea
n^{s}_{b}&=&\frac{m^{*}_{b}}{2\pi^{2}} \left[E_{Fb}k_{Fb}-{m}^{*2}_{b}\ln
\frac{k_{Fb}+E_{Fb}}{m^{*}_{b}} \right] \ , \nonumber \\
n_{b}&=&\frac{k_{Fb}^{3}}{3\pi^{2}} \ , \nonumber \\
n_{l}&=&\frac{k_{Fl}^{3}}{3\pi^{2}} \ .
\eea

With the above ingredients, one can compute the energy density and the pressure of the system.  
The energy density is given by 
\bea
\varepsilon&=&\sum_{b} \varepsilon_{b}+\sum_{l}\varepsilon_{l} \nonumber \\ &+&\frac{1}
{2}m^{2}_{\sigma} \bar \sigma^{2}+\frac{1}{2}m^{2}_{\omega} \bar \omega^{2}+\frac{1}{2}m^{2}_{\rho} \bar \rho^{2} +\frac{1}{2}m^{2}_{\phi} \bar \phi^{2}\cr
&+& \frac{\kappa}{3 !} (g_{\sigma} \bar \sigma)^ 3+ \frac{\lambda}{4 !} (g_{\sigma} \bar \sigma)^ 4 \cr
&+& \frac{\zeta}{8} (g_{\omega} \bar \omega)^4 + 3 \Lambda_{\omega} (g_{\rho} g_{\omega} \bar \rho \, \bar \omega) ^2 ,
\label{enerdens}
\eea 
where the energy densities of baryons and leptons take the expressions
\bea
\varepsilon_{b}&\!\!\!=\!\!\!&\frac{1}{8\pi^ {2}}\bigg[k_{Fb}E_{Fb}^{3}+k_{Fb}^{3} E_{Fb} - {m}^{*4}_{b}\ln\frac{k_{Fb}+E_{Fb}}{{m}^{*}_{b}}  \bigg] , \nonumber \\ 
\varepsilon_{l}&\!\!\!=\!\!\!&\frac{1}{8\pi^ {2}}\bigg[k_{Fl}E_{Fl}^{3}+k_{Fl}^{3} E_{Fl} - m^{4}_{l}\ln\frac{k_{Fl}+E_{Fl}}{m_{l}}  \bigg] \!.
%
\eea
We note that in obtaining Eq.~(\ref{enerdens}) for the energy density, the 
equations of motion (\ref{eqphi}) were used to rewrite the contribution to  
$\varepsilon$ of $\sum_{b} ( g_{\omega b} \bar\omega n_b + g_{\rho b} \bar\rho 
I_{3 b} n_b + g_{\phi b} \bar\phi n_b)$.
Finally, the pressure can be computed using the thermodynamic relation
\beq
P=\sum_{i}\mu_{i}n_{i}-\varepsilon ,
\label{press}
\eeq
where the baryonic and leptonic chemical potentials are given by 
\bea
\mu_{b}&=& E_{Fb}+g_{\omega  b} \, \bar \omega+g_{\rho b} \,I_{3 b}   \,\bar \rho +g_{\phi b} \,\bar \phi \ , \nonumber \\
\mu_{l} &=& E_{Fl}.
\eea

The cores of neutron stars harbor globally neutral matter that is in $\beta$-equilibrium.
Therefore, the chemical potentials and the number densities of the different particles in a neutron 
star core are related by the conditions
\bea
&&\mu_i=b_i \mu_n- q_i \mu_e \ ,\nonumber \\
&&0=\sum_{b,l} q_i \, n_i \ ,  \nonumber \\
&& n=\sum_{b} n_i \ ,
\label{beta-eq}
\eea
where $b_i$ and $q_i$ denote, respectively, the baryon number and the charge of the particle $i$. These relations, 
the Dirac equations (\ref{MFlep}) for the baryons and leptons, and the field equations
(\ref{eqphi}) for the mesonic fields $\sigma$, $\omega$, 
$\rho$ and $\phi$, are to be solved self-consistently for a given total baryon density $n$. 
Once the chemical potential and the density of each species have been obtained at the given
$n$, one can determine the energy density and pressure of the neutron star matter for each density.

\section{Models for the equation of state}
\label{sec:eos}

\begin{table*}[t!]
\caption{Parameters of the models FSU2R and FSU2H of this work. The mass of the nucleon is $m_N= 939$ MeV.}
\centering
\begin{tabular*}{\textwidth}{ccccccccccc}
\hline\hline
Model & $m_{\sigma}$ & $m_{\omega}$ & $m_{\rho}$ & $g_{\sigma N}^2$ & $g_{\omega N}^2$ & $g_{\rho N}^2$ & $\kappa$ & $\lambda$ & $\zeta$ & $\Lambda_{\omega}$ \\
      & (MeV) & (MeV) & (MeV) & & &  & & &  &  \\
\hline
%
%
FSU2R & 497.479 & 782.500 & 763.000 & 107.5751 & 182.3949 & 206.4260 & 3.0911 & $-$0.001680 & 0.024  & 0.045 \\
FSU2H & 497.479 & 782.500 & 763.000 & 102.7200 & 169.5315 & 197.2692 & 4.0014 & $-$0.013298 & 0.008  & 0.045 \\
\hline\hline
\end{tabular*}
\label{t-parameters}
\end{table*}

\begin{table*}[t!]
\caption{Properties at saturation of the models FSU2R and FSU2H of this work.
We show the saturation density 
($n_0$), energy per particle ($E/A$), compressibility ($K$), and effective nucleon 
mass ($m_N^*/m_N$) in symmetric nuclear matter, as well as the symmetry energy ($E_{\rm 
sym}$), slope of the symmetry energy ($L$), curvature of the symmetry energy ($K_{\rm 
sym}$), and pressure of pure neutron matter ($P_{\rm PNM}$) at $n_0$.}
\centering
\begin{tabular*}{\textwidth}{c\x cccccccc}
\hline\hline
Model  & $n_0$ & $E/A$ & $K$ & $m_N^*/m_N$ & $E_{\rm sym}(n_0)$ & $L$ & $K_{\rm sym}$ & $P_{\rm PNM}(n_0)$ \\ 
       & $({\rm fm}^{-3})$ & $({\rm MeV})$ & $({\rm MeV})$ & & $({\rm MeV})$ & (MeV) & $({\rm MeV})$ & $({\rm MeV \, fm}^{-3})$ \\ 
\hline
%
FSU2R & 0.1505 & $-$16.28 & 238.0 & 0.593 & 30.7 & 46.9  & 55.7    & 2.44 \\
FSU2H & 0.1505 & $-$16.28 & 238.0 & 0.593 & 30.5 & 44.5  & 86.7    & 2.30 \\
\hline\hline
\end{tabular*}
\label{t-props}
\end{table*}

From the Lagrangian (\ref{lagran}), in \citep{Tolos:2016hhl} we formulated the models 
FSU2R (with nucleons) and FSU2H (with nucleons and hyperons), with a 
motivation for accomodating massive enough stars and the new astrophysical 
measurements of small stellar radii 
within a self-consistent microscopic theory of the EoS for the core of neutron stars. 
Note that this type of approach is different \mbox{from---but complementary to---the} 
methods where the astrophysical and nuclear observables are mapped onto the EoS through 
piecewise parametrizations of the EoS 
\citep{Raithel:2016bux,Lattimer:2015nhk,Ozel:2016oaf}.
To build our models we started from the nucleonic FSU2 model of \citep{Chen:2014sca} 
that reproduces heavy neutron star masses but was not constrained to radii. The 
condition of small stellar radii imposed a soft nuclear symmetry energy in the theory. 
We showed that the resulting FSU2R and FSU2H models, besides the mentioned 
astrophysical constraints, can successfully describe the properties of finite 
nuclei and conform to the constraints on the nuclear EoS from kaon production and 
collective flow in HICs~\citep{Fuchs:2000kp,Lynch:2009vc,Danielewicz:2002pu}.

In the present work we start by introducing a modification of the parameters of 
our models FSU2R and FSU2H of \citep{Tolos:2016hhl} in order to refine the behavior 
of the EoS of pure neutron matter (PNM) in the region of subsaturation densities. 
We report the new version of the parameters in Table~\ref{t-parameters} 
(it should be mentioned that the form of the equations of motion remains the same 
irrespective of the specific values of the coupling constants). We 
have changed the value of the quartic isovector-vector coupling $\Lambda_{\omega}$ of 
FSU2R and FSU2H from 0.05 in \citep{Tolos:2016hhl} to 0.045. This has been done because
it results in a symmetry energy that is a little stiffer than before and avoids a previous 
instability in the EoS of PNM for low subsaturation densities. As $\Lambda_{\omega}$ 
has been changed, we have refitted accordingly the value of the coupling $g_{\rho N}^2$ 
between the $\rho$-meson and the nucleons to obtain the same good reproduction of 
binding energies and charge radii of finite nuclei as in \citep{Tolos:2016hhl}. The 
values of the other parameters of FSU2R and FSU2H are the same of 
\citep{Tolos:2016hhl}.
Owing to the fact the $\Lambda_{\omega}$ and $g_{\rho N}^2$ couplings only 
contribute in neutron-rich matter, the EoS of symmetric nuclear matter (SNM), composed 
of the same number of protons and neutrons, is identical to that of our models FSU2R 
and FSU2H in \citep{Tolos:2016hhl}.

We collect in Table~\ref{t-props} a few characteristic isoscalar and isovector
properties at the nuclear matter saturation density $n_0$ for the present version of 
our models.  
In the work \citep{Fortin:2014mya}, the authors derived the constraint $1.7 
\lesssim P(n_0) \lesssim 2.8 \mbox{ MeV fm}^{-3}$ for the pressure of neutron star 
matter at saturation density. They deduced this constraint from the results of the 
microscopic calculations of PNM performed by \citep{Hebeler:2013nza} using chiral 
two-nucleon and three-nucleon interactions, which are in good agreement with the 
results by \citep{Gandolfi:2011xu} from Quantum Monte Carlo calculations with the 
Argonne $v_{18}$ nucleon-nucleon potential plus three-nucleon forces.
A narrower range $2.3 \lesssim P(n_0) \lesssim 2.6 \mbox{ MeV fm}^{-3}$ was  
estimated more recently by \citep{Hagen:2015yea} from ab initio calculations 
of nuclear systems with chiral interactions. 
While our models FSU2R and FSU2H of \citep{Tolos:2016hhl}, with PNM pressures at 
saturation of 2.27 and 2.06 MeV fm$^{-3}$, fulfill the constraint of 
\citep{Fortin:2014mya}, they are somewhat below the constraint of \citep{Hagen:2015yea}.
Now, with the new parametrization of our models, we are able to obtain PNM pressures 
at saturation density of 2.44 MeV fm$^{-3}$ in FSU2R and of 2.30 MeV fm$^{-3}$ in FSU2H 
(see Table~\ref{t-props}) that are consistent with both the predictions from chiral 
forces derived by \citep{Fortin:2014mya} and \citep{Hagen:2015yea}.

The EoS of PNM of the present FSU2R and FSU2H models differs from the results we showed 
in \citep{Tolos:2016hhl} almost only for densities in the low-density region $n 
\lesssim n_0$, where the pressure of PNM is a little higher now. However, above 
saturation density, the pressures of our current parameters and those of 
\citep{Tolos:2016hhl} are very similar.
Consequently, compared with \citep{Tolos:2016hhl}, one may anticipate that the 
predictions for masses and radii of neutron stars will not be drastically affected. 

\begin{figure}[t]
\begin{center}
\includegraphics[width=0.48\textwidth]{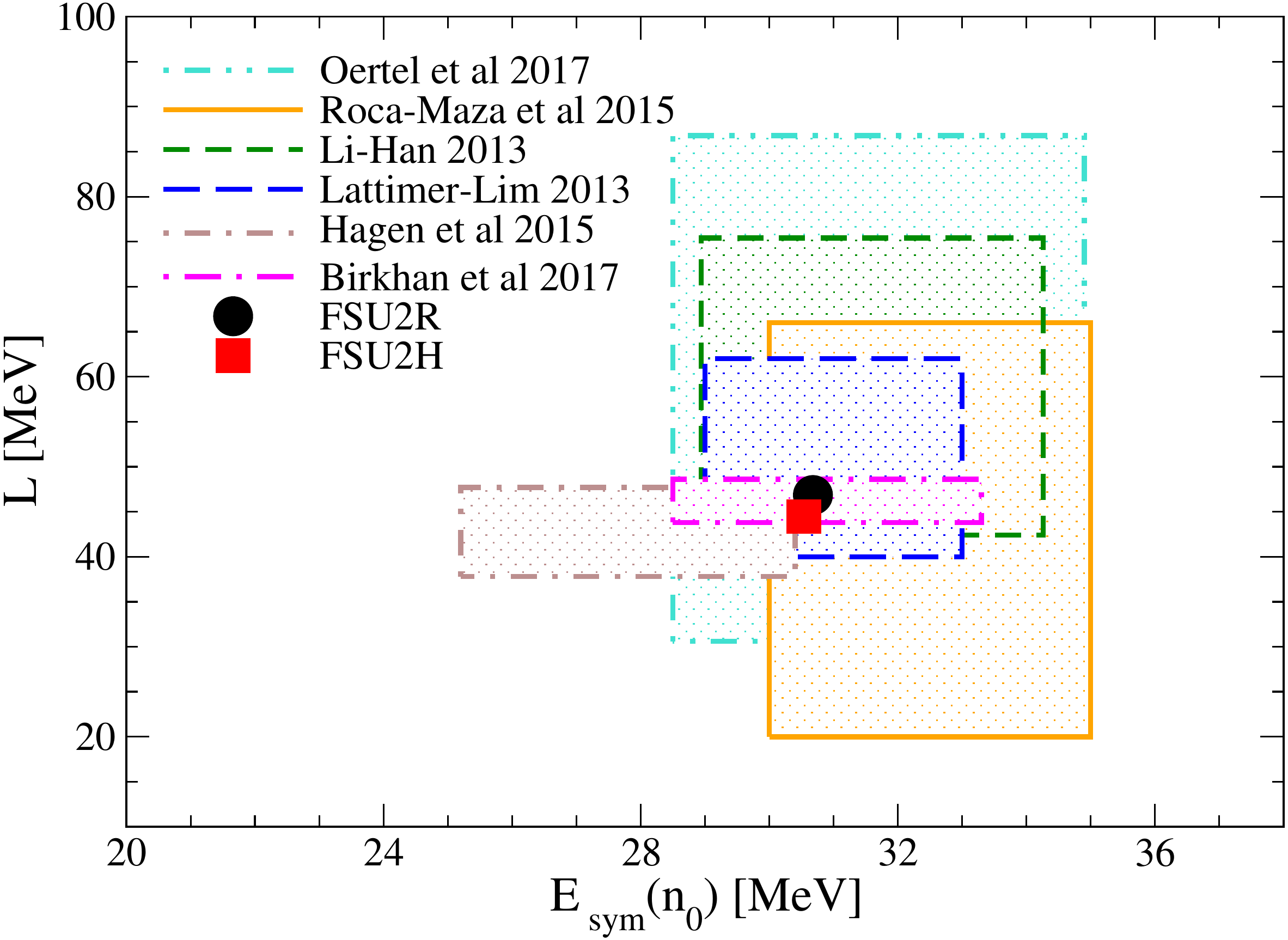}
\caption{Slope of the symmetry energy ($L$) versus symmetry energy ($E_{\rm sym}(n_0)$) 
at the nuclear matter saturation density for the models FSU2R and FSU2H 
discussed in text. The shaded regions depict the determinations from
\citep{Li:2013ola,Lattimer:2012xj,Roca-Maza:2015eza,Hagen:2015yea,Oertel:2016bki,Birkhan:2016qkr}.}
\label{fig:esym-l}
\end{center}
\end{figure}

The slope parameter $L$ of the symmetry energy, i.e., 
$\displaystyle L = 3 n_0 \Big( \frac{\partial E_{\rm sym}(n)}{\partial n} \Big)_{n_0}$, 
has become a standard reference in the literature for characterizing the stiffness of 
the change of the nuclear symmetry energy $E_{\rm sym}(n)$ with density.
In our original version of the FSU2R and FSU2H models shown in \citep{Tolos:2016hhl}, 
the value of the symmetry energy at saturation density was $E_{\rm sym}(n_0)= 30.2$~MeV 
in both models, while the slope parameter was $L= 44.3$~MeV in FSU2R and $L= 41$~MeV in 
FSU2H. In the updated version of FSU2R and FSU2H of the present work, these properties 
become $E_{\rm sym}(n_0)= 30.7$~MeV and $L= 46.9$~MeV in FSU2R and $E_{\rm sym}(n_0)= 
30.5$~MeV and $L= 44.5$~MeV in FSU2H (see Table~\ref{t-props}). 
These values suggest a relatively soft nuclear symmetry energy. 
We have plotted in Fig.~\ref{fig:esym-l} the ranges for $E_{\rm sym}(n_0)$ and 
$L$ that have been estimated in several recent works through the analysis of a variety 
of nuclear data from terrestrial experiments, astrophysical observations, and 
theoretical calculations 
\citep{Li:2013ola,Lattimer:2012xj,Roca-Maza:2015eza,Hagen:2015yea,Oertel:2016bki,Birkhan:2016qkr}. 
It can be seen that the predictions of FSU2R and FSU2H have an overlap with the 
majority of these ranges. We would like to remark that this is an {\sl a posteriori} 
result, because the predicted values of $E_{\rm sym}(n_0)$ and $L$ are the consequence 
\citep{Tolos:2016hhl} of having adjusted the FSU2R and FSU2H parameter sets to 
reproduce neutron star radii of about 13 km, without sacrificing maximum  
masses of $2 M_\odot$ nor the description of binding energies and charge radii of 
atomic nuclei.
Hence, we interpret the reasonable agreement of our results with the multiple 
constraints in Fig.~\ref{fig:esym-l} as hinting at the plausibility of the 
existence of neutron stars with relatively small radii.

The neutron matter EoS is also strongly related with the neutron 
distribution in atomic nuclei. Models with softer symmetry energies produce a thinner 
neutron skin $\Delta r_{np}$ (difference between the rms radii of the neutron and 
proton density distributions) in nuclei \citep{Brown2000,Horowitz:2000xj}. 
Unfortunately, neutron densities and neutron radii are poorly 
known to date because the distribution of neutrons in a nucleus is hard to measure. Our 
present FSU2R and FSU2H models predict a neutron skin thickness of 0.15 fm in the 
neutron-rich nucleus $^{208}$Pb. This prediction is compatible with the range 
$0.13 \lesssim \Delta r_{np} \lesssim 0.19$ fm for $^{208}$Pb extracted in  
\citep{Roca-Maza:2015eza} from measurements of the electric dipole polarizability of
nuclei, the value $\Delta r_{np} = 0.15 \pm 0.03$ fm determined from coherent pion 
photoproduction in $^{208}$Pb at the MAMI facility \citep{Tarbert:2013jze}, and the
value $\Delta r_{np} = 0.302 \pm 0.177$ fm from parity violating electron scattering on 
$^{208}$Pb performed at JLab \citep{Abrahamya12,Horowitz:2012tj}.\footnote{We note that while  
experimental data are always provided with the associated error bars, theoretical models like ours, after the values of the 
coupling constants have been specified, make ``exact'' predictions with no error bars. In the future,  
it will be worth estimating error bars on our theoretical results, following recent initiatives to 
assess statistical errors and error propagation in nuclear functionals \citep{Dobaczewski:2014jga,Chen:2014sca}.}
In the case of the lighter nucleus $^{48}$Ca, we find a neutron radius of 3.55 fm with 
FSU2R and of 3.57 fm with FSU2H, and a neutron skin of 0.166 fm with both 
models. The prediction is in good accord with the ranges 3.47--3.60 fm for the neutron 
radius and 0.12--0.15 fm for the neutron skin of $^{48}$Ca obtained in
\citep{Hagen:2015yea} through ab initio calculations of the neutron distribution of 
$^{48}$Ca using nuclear interactions derived from chiral effective field theory;
it also is in accord with the neutron skin of 0.14--0.20 fm for $^{48}$Ca found 
from the new measurement of the electric dipole polarizability in $^{48}$Ca \citep{Birkhan:2016qkr}.
Altogether, it appears that the properties of the symmetry energy of the proposed 
models for the EoS, which are motivated by reproducing small neutron star radii (see 
next section), are compatible within uncertainties with different empirical and 
theoretical extractions of these properties.


\section{Stellar properties}
\label{sec:stellar}

Having access to the pressure and energy density of matter, we can compute the 
properties of neutron stars by solving the Tolman-Oppenheimer-Volkoff (TOV) equations 
\citep{Oppenheimer:1939ne}. For static and spherically-symmetric stars, the TOV 
equations read as
\bea
\frac{dP}{dr}&=&-\frac{G}{r^2}\left(\varepsilon+P\right)
\left(m+4\pi r^3 P\right)\left(1-\frac{2Gm}{r}\right)^{-1} , \nonumber \\
\frac{dm}{dr}&=&4 \pi r^2 \varepsilon ,
\label{tov}
\eea
where $r$ is the radial coordinate, $m$ is the mass enclosed by a radius $r$, and $G$ 
is the gravitational constant. For a given central density, the integration of these 
equations provides the corresponding mass and radius of the star. 
By repeating the calculation for different central densities, the 
mass-radius (M-R) relation of neutron stars can be obtained.

Indeed, to solve the TOV equations for a neutron star we need the EoS of matter
over a wide range of densities from the center to the surface of the star.
The structure of a neutron star is such that the heavy liquid core is surrounded by a 
thin solid crust \citep{Shapiro:1983du,Haensel:2007yy}.
The transition from the core to the crust occurs when the density of matter 
becomes lower than approximately $1.5 \times 10^{14}$~g/cm$^3$. Below this density, 
matter ceases to exist in a homogeneous liquid phase because it is favorable 
that the protons concentrate with neutrons in nuclear clusters, which 
arrange themselves in a crystal lattice in order to minimize the Coulomb repulsion 
among them \citep{Baym:1971ax,Baym:1971pw,Shapiro:1983du,Haensel:2007yy}.
In the inner layers of the crust, the nuclear clusters are beyond the neutron drip 
point and the lattice is permeated by a gas of free neutrons in addition to the electron 
gas, whereas in the outer crust the nuclear clusters are neutron-rich nuclei below the 
neutron drip point, embedded in the electron gas.
We have solved the TOV equations using the FSU2R and FSU2H models for 
the EoS of the uniform matter of the liquid core of the star for densities above 0.09~fm$^{-3}$ 
($\approx 1.5 \times 10^{14}$~g/cm$^3$), under the 
conditions of $\beta$-equilibrium and global charge neutrality expressed in
Eq.~(\ref{beta-eq}) of Sec.~\ref{sec:formalism}.
At the densities of the crust, in the absence of calculations with our models of 
the complex structures that can populate this region of the star, we have used 
the EoS for the crust of neutron stars that has recently been derived from calculations 
based on the Brueckner theory in \citep{Sharma:2015bna}.


The FSU2R model applies to nucleonic cores of neutron stars, i.e., when 
the whole stellar core is assumed to consist of neutrons, protons, electrons, and muons 
(`$npe\mu$' matter). In the dense inner region of a neutron star core, however, 
the chemical potential may become so high that matter will be able to undergo a 
transition to other states of the low-lying octet of baryons, with hyperons appearing in 
the composition (`$npY\!e\mu$' matter). Thus, we have devised the 
FSU2H model to allow for the presence of hyperons in the star interior  
\citep{Tolos:2016hhl}. In consequence, in the case of the FSU2H EoS, besides the 
nucleon and meson couplings shown in Table~\ref{t-parameters}, we have considered the 
complete octet of baryons in the Lagrangian density (\ref{lagran}). We have fixed 
the corresponding hyperon couplings from SU(3) flavor symmetry and from information on 
hyperon optical potentials in hypernuclei. 
We leave for the next section the discussion of the determination of the hyperon 
couplings and the analysis of the influence of the uncertainties associated with these 
couplings. Here, we focus on the results for the masses and sizes of neutron stars from 
the FSU2R nucleonic EoS and from the FSU2H hyperonic EoS with our baseline values for 
the hyperon couplings (given in Sec.~\ref{sec:hyperon}).

\begin{figure}[t!]
\begin{center}
\includegraphics[width=0.48\textwidth]{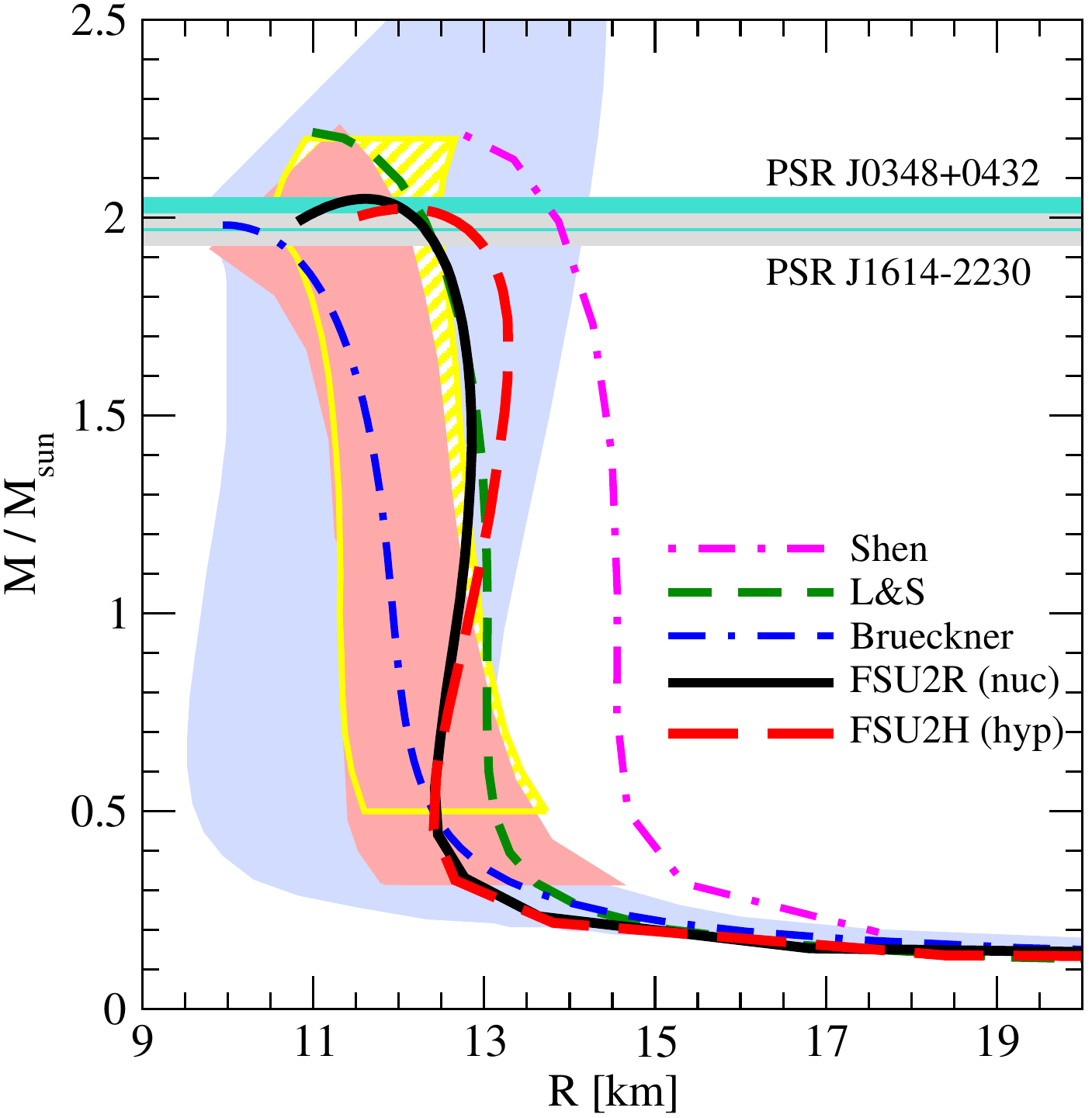}
\caption{Mass versus radius for neutron stars from the models FSU2R and FSU2H of this 
work and from some models from the literature (Shen: \citep{Shen:1998gq}, L\&S: 
\citep{ls91}, Brueckner: \citep{Sharma:2015bna}). The thin horizontal bands indicate 
the heaviest observed masses $M=1.97 \pm 0.04 M_\odot$ \citep{Demorest:2010bx} and 
$M=2.01 \pm 0.04 M_\odot$ \citep{Antoniadis:2013pzd}. 
The vertical blue band at the back depicts the M-R region constrained in 
\citep{Hebeler:2013nza} from chiral nuclear interactions up to $n=1.1n_0$ 
and the conditions of $M_{\rm max}> 1.97 M_\odot$ and causality.
The vertical red band at the front shows the M-R region derived from five 
quiescent low-mass X-ray binaries and five photospheric radius expansion X-ray bursters 
after a Bayesian analysis \citep{Lattimer:2014sga}.
The vertical striped yellow band is the M-R constraint derived from the cooling tails 
of type-I X-ray bursts in three low-mass X-ray binaries and a Bayesian analysis in 
\citep{Nattila:2015jra} (model A of the paper).}
\label{fig:mass-radius-new}
\end{center}
\end{figure}

We display the results for the relation between mass and radius of neutron stars in 
Fig.~\ref{fig:mass-radius-new}. A few data on the maximum mass configuration and the 
$1.5 M_\odot$ configuration from FSU2R and FSU2H are presented in 
Table~\ref{tab:starprops}.
For completeness, in Fig.~\ref{fig:mass-radius-new}, besides the curves of FSU2R and 
FSU2H, we also plot the M-R relations of two popular EoSs widely used in astrophysical 
calculations. They correspond to the Shen et al.\ EoS based on the relativistic TM1 
nuclear mean field model \citep{Shen:1998gq} and to the Lattimer--Swesty EoS based on a 
non-relativistic Skyrme nuclear force (in its Ska version) \citep{ls91}. Also shown is 
the result of the recent EoS from the Brueckner theory with the Argonne $v_{18}$ 
potential plus three-body forces computed with the Urbana model \citep{Sharma:2015bna}. 
These additional EoSs are all non-hyperonic. We have included in the same 
Fig.~\ref{fig:mass-radius-new} a few recent astrophysical determinations of neutron star
mass-radius limits.

\begin{table*}[t!]
\caption{Properties of the maximum mass and $1.5 M_\odot$ configurations for nucleonic ($npe\mu$) 
neutron stars calculated with the FSU2R EoS and for hyperonic ($npY\!e\mu$) 
neutron stars calculated with the FSU2H EoS. From top to bottom, mass, radius, compactness parameter $G M / R c^2$,
surface gravitational red shift $z_{\rm surf}= (1- 2 G M / R c^2)^{-1/2}-1$, and the values 
of the number density, pressure, and mass-energy density at the center of the star.}
\centering
%
\begin{tabular*}{\textwidth}{\x lcccc}
\hline\hline
 & \multicolumn{2}{c}{$M_\text{max}$ configuration} & \multicolumn{2}{c}{$1.5 M_\odot$ configuration} \\
   \cline{2-3}                            \cline{4-5}
 & FSU2R & FSU2H & FSU2R & FSU2H \\
 & (nuc) & (hyp) & (nuc) & (hyp) \\
\hline
$M / M_\odot$        & 2.05 & 2.02 & 1.50 & 1.50 \\
$R$ (km)             & 11.6 & 12.1 & 12.8 & 13.2 \\
$G M / R c^2$        & 0.26 & 0.25 & 0.17 & 0.17 \\
$z_{\rm surf}$       & 0.45 & 0.40 & 0.24 & 0.23  \\
%
%
%
$n_c/n_0$           & 6.3  & 5.8  & 2.7  & 2.3 \\
$P_c$ ($10^{15}$\,g\,cm$^{-3}$)             & 0.62  & 0.46 & 0.11 & 0.09 \\
$\varepsilon_c$ \,($10^{15}$\,g\,cm$^{-3}$) & 2.08  & 1.80 & 0.75 & 0.63 \\
%
%
\hline\hline
\end{tabular*}
\label{tab:starprops}
\end{table*}

The M-R curve from each EoS exhibits a maximum mass, beyond which the star would become 
unstable against collapse into a black hole.
The heaviest known masses of neutron stars are $M=1.97 \pm 0.04 M_\odot$ in the  
PSR J1614--2230 pulsar \citep{Demorest:2010bx} and $M=2.01 \pm 0.04 M_\odot$ in the 
PSR J0348+0432 pulsar \citep{Antoniadis:2013pzd}. We depict them by the horizontal 
bands in Fig.~\ref{fig:mass-radius-new}. Both the nucleonic FSU2R EoS and the hyperonic 
FSU2H EoS are able to provide maximum masses fulfilling the $\approx 2 M_\odot$ 
observational limit, as well as the other EoSs shown in the same figure. We note that 
FSU2R reaches the maximum mass with a fairly compact stellar radius of 11.6~km (see 
Table~\ref{tab:starprops}). For canonical neutron stars with masses of $1.4 
M_\odot$--$1.5 M_\odot$, FSU2R predicts a radius of 12.8~km. 
The recent astrophysical determinations of neutron star radii from quiescent low-mass 
X-ray binaries in globular clusters and X-ray bursters seem to point in this direction
\citep{Guillot:2013wu,Guillot:2014lla,Guver:2013xa,Heinke:2014xaa,Lattimer:2013hma,
Lattimer:2014sga,Ozel:2015fia}. Although these determinations are indirect and depend 
on stellar atmosphere models, they overall converge in favoring small neutron star 
radii in the range of about \mbox{9--13~km} \citep{Lattimer:2015nhk,Ozel:2016oaf}. 
An accurate radius measurement by new observatories such as NICER
\citep{2014SPIE.9144E..20A}, which has begun operating aboard the International Space 
Station in June 2017, would represent a major step forward to corroborate or modify these 
expectations.

The compromise between having large maximum masses and small radii for canonical 
neutron stars is a challenging constraint that rules out a large number of 
theoretical EoSs \citep{Lattimer:2015nhk,Ozel:2016oaf,Oertel:2016bki}.
This follows from the fact that the pressure of the high-density EoS must be hard 
enough to sustain massive stars, whereas the pressure at 1--2 times the nuclear 
saturation density $n_0$ must be, in contrast, effectively soft in order to produce 
small radii for canonical mass stars. Given that the pressure of neutron star matter in 
the vicinity of $n_0$ is basically governed by the nuclear symmetry energy, the 
challenge is particularly acute in relativistic field theoretical models because the 
relativistic models usually have stiff symmetry energies. As we have demonstrated with 
FSU2R, it is possible to obtain parametrizations of the considered relativistic 
Lagrangian that meet both large stellar masses \mbox{and---on} condition of a 
soft symmetry \mbox{energy---radii} smaller than $\sim\! 13$ km for $M \gtrsim 1.4 
M_\odot$, and that still provide an excellent reproduction of the binding energies and 
charge radii of finite nuclei \citep{Tolos:2016hhl}. 
There are some other parametrizations in the frame of the relativistic field 
theory that support these findings, such as the recent RMF012 and RMF016 models 
of \citep{Chen:2014mza,Chen:2015zpa}. Indeed, the accurately calibrated RMF016 model 
produces neutron stars of $2 M_\odot$ and gives radii of 13 km for stars of $1.4 
M_\odot$ \citep{Chen:2014mza,Chen:2015zpa}, in keeping with the predictions of our 
FSU2R EoS.

When we allow for the appearance of hyperons in the neutron star core with the FSU2R 
model, the maximum mass of the star experiences a reduction of the order of 15\%, due 
to the expected softening of the EoS, and then, with a maximum mass of $1.77 M_\odot$, 
it falls short of the $2 M_\odot$ limit.
In an effort to shed some light on the question whether with exotic degrees of freedom 
in the core, the star can satisfy the targets of $2 M_\odot$ maximum mass and small 
radius at canonical mass, we have developed the hyperonic model FSU2H. In FSU2H, we 
essentially have stiffened further the nucleonic pressure above twice the saturation 
density, i.e., around the onset of appearance of hyperons. This 
comes at the price of a certain overpressure in symmetric nuclear matter for densities 
$n \gtrsim 2 n_0$ when we compare it with the constraints deduced from the modeling of 
collective flow in HICs \citep{Danielewicz:2002pu}, cf.\ Fig.~1 of 
\citep{Tolos:2016hhl}. Yet the pressure of FSU2H in pure neutron matter, shown 
also in Fig.~1 of \citep{Tolos:2016hhl}, fits within the projected region from the 
collective flow studies. Given that the $\beta$-equilibrated neutron-star matter is 
highly asymmetric, we consider this model as sufficiently realistic for describing 
neutron stars. The determination of narrower constraints on the EoS of PNM at several 
times $n_0$ from HIC experiments \citep{Russotto:2016ucm} in the future should be of 
great help in this regard.

It can be observed in Fig.~\ref{fig:mass-radius-new} that the FSU2H model with hyperons 
produces a comparable M-R relation to FSU2R and satisfies, as mentioned,  
the observational limit of $2 M_\odot$. With respect to FSU2R, in FSU2H the 
size of the radii has increased by 0.2--0.5~km for neutron stars heavier than $1 
M_\odot$, expectedly, from the stiffer pressure of the nucleonic sector above twice 
the saturation density. The onset of hyperons occurs at a baryon density of 0.33 
fm$^{-3}$, or $2.2 n_0$. The maximum mass of $2.02 M_\odot$ calculated with FSU2H is 
characterized by a radius of 12.1 km (see Table~\ref{tab:starprops}). For $1.5 M_\odot$ 
stars, the hyperonic FSU2H EoS predicts radii of 13.2~km, which, although on the upper 
edge, are still compatible with the recent astrophysical indications of neutron star 
radii of about 9--13~km \citep{Lattimer:2015nhk,Ozel:2016oaf}. 
The numerical results for the EoS and M-R relation of the FSU2R and FSU2H models 
are tabulated in the Appendix.

In closing this section, we ought to mention that the results for stellar radii of our 
EoSs have been possible while obtaining, within the same models, a realistic 
reproduction of the properties of atomic nuclei and of several other constraints.
It seems unlikely that one may be able to account for significantly smaller neutron 
star radii in the theory considered here without abandoning the physical region 
of parameters. Hence, a discovery of even smaller stellar radii could provide evidence 
in favor of a phase transition to other degrees of freedom in neutron star
interiors~\citep{Dexheimer:2015qha}.


\section{Impact of uncertainties in the hyperon couplings}
\label{sec:hyperon}

We next discuss the determination of the values of the hyperon couplings in our FSU2H 
EoS and estimate the influence that the uncertainties in these couplings may have on 
the predictions for neutron star masses and radii. 

We recall that the potential felt by a hyperon $i$ in $j$-particle matter is given by
\bea
&&U_i^{(j)}(n_j) = \noindent \nonumber \\
&&- g_{\sigma i} \, \bar \sigma^{(j)} + g_{\omega i} \,  \bar \omega^{(j)} +  g_{\rho i}  \, I_{3 i} \, \bar \rho^{(j)} + g_{\phi i} \, \bar \phi^{(j)},  \  \ \ \ \ \
\label{Ypot}
\eea
in our model, 
where $ \bar \sigma^{(j)}$, $\bar \omega^{(j)}$, $\bar \rho^{(j)}$ and $\bar \phi^{(j)}$ are the meson field values in $j$-particle matter while $I_{3i}$ denotes the third component of the isospin operator. Flavor SU(3) symmetry, the vector dominance model and ideal mixing for the physical $\omega$ and $\phi$ mesons, permit relating the couplings between the hyperons and the vector mesons to the nucleon couplings $g_{\omega N}$ and $g_{\rho N}$  \citep{Schaffner:1995th,Banik:2014qja,Miyatsu:2013hea,Weissenborn:2011ut,Colucci:2013pya,Tolos:2016hhl}, according to the ratios 
\begin{eqnarray}
g_{\omega \Lambda}:g_{\omega \Sigma}:g_{\omega \Xi}:g_{\omega N}&=&\frac{2}{3}:\frac{2}{3}:\frac{1}{3}:1 \nonumber \\
g_{\rho \Lambda}:g_{\rho \Sigma}:g_{\rho \Xi}:g_{\rho N}&=&0:1:1:1 \nonumber \\
g_{\phi \Lambda}: g_{\phi\Sigma}: g_{\phi\Xi}:g_{\omega N}&=& -\frac{\sqrt{2}}{3}: -\frac{\sqrt{2}}{3}:  -\frac{2\sqrt{2}}{3}: 1 , \ \ \ \ \ \ \ 
\label{eq:couplings}
\end{eqnarray}
and noting that $g_{\phi N}=0$. We reduce by 20\% the coupling of the  $\Lambda$ hyperon to the $\phi$ meson in order to obtain a $\Lambda\Lambda$ bond energy in $\Lambda$ matter at a density $n_\Lambda \simeq n_0/5$ of $\Delta B_{\Lambda\Lambda} (n_0/5) = 0.67$ MeV, thereby reproducing the value extracted from the 
$^6_{\Lambda\Lambda} {\rm He}$ double $\Lambda$ hypernucleus, also known as the Nagara event \citep{Takahashi:2001nm, Ahn:2013poa}. 

The coupling of each hyperon to the scalar $\sigma$ meson field is left as a free parameter to be adjusted to reproduce the hyperon potential in SNM, derived from hypernuclear data. It is well known that a Woods-Saxon type potential of depth $U_\Lambda^{(N)}(n_0) \sim -28$ MeV reproduces the bulk of $\Lambda$ hypernuclei binding energies \citep{Millener:1988hp}. As for the $\Sigma$ hyperon, a moderate repulsive potential could be extracted from analyses of  $(\pi^-, K^+)$ reactions off nuclei  \citep{Noumi:2001tx} done in \citep{Harada:2006yj,Kohno:2006iq}. Fits to $\Sigma^-$ atomic data \citep{Friedman:2007zza} also point towards a transition from an attractive $\Sigma$-nucleus potential at the surface to a repulsive one inside the nucleus, the size of  the repulsion not being well determined. The potential felt by a $\Xi$ hyperon in SNM is also quite uncertain. Old emulsion data indicate sizable attractive values of 
 around  $U_\Xi^{(N)}(n_0) = - 24 \pm 4$ MeV \citep{Dover:1982ng}, while the
analyses of the $(K ^-, K^+)$ reaction on a $^{12}$C target suggest a milder attraction \citep{Fukuda:1998bi,Khaustov:1999bz}. Taking these experimental uncertainties into account, we allow the hyperon potentials in SNM to take the following range of values:
 \begin{eqnarray}
U_{\Lambda}^{(N)}(n_0)&=&-28~{\rm MeV} \nonumber \\
U_{\Sigma}^{(N)}(n_0)&=&0~{\rm to}~ 30~{\rm MeV} \nonumber \\
U_{\Xi}^{(N)}(n_0)&=&-18~{\rm to}~0~{\rm MeV}  \ ,
\label{eq:pots}
\end{eqnarray}
Note that we only consider uncertainties for the $\Sigma$ and $\Xi$ potentials, given the consensus on the $\Lambda$ potential at saturation. The range of values for the hyperon potentials in SNM give rise to the following range for the hyperon-$\sigma$ couplings:
\begin{eqnarray}
g_{\sigma \Lambda}/g_{\sigma N}&=& 0.611  \nonumber \\
g_{\sigma  \Sigma}/g_{\sigma N}&=& 0.467 -  0.541  \nonumber \\
g_{\sigma  \Xi}/g_{\sigma N} &=& 0.271 - 0.316   \ ,
 \label{eq:sigma} 
\end{eqnarray}
where the lower values correspond to the most repulsive situation ($U_\Sigma^{(N)}(n_0) =30$ MeV, $U_\Xi^{(N)}(n_0)=0$ MeV) and the upper ones to the most attractive one ($U_\Sigma^{(N)}(n_0) =0$ MeV, $U_\Xi^{(N)}(n_0)=-18$ MeV). 
In our baseline FSU2H model used in the calculations of Sec.~\ref{sec:stellar} we have adopted the values $U_{\Lambda}^{(N)}(n_0)=-28$ MeV, $U_\Sigma^{(N)}(n_0) = 30$ MeV and $U_\Xi^{(N)}(n_0)= -18$ MeV, which lead to the couplings $g_{\sigma \Lambda}= 0.611 g_{\sigma N}$, $g_{\sigma \Sigma}=  0.467 g_{\sigma N}$ and $g_{\sigma \Xi}= 0.316 g_{\sigma N}$.  

\begin{figure}[t!]
\begin{center}
\includegraphics[width=0.48\textwidth]{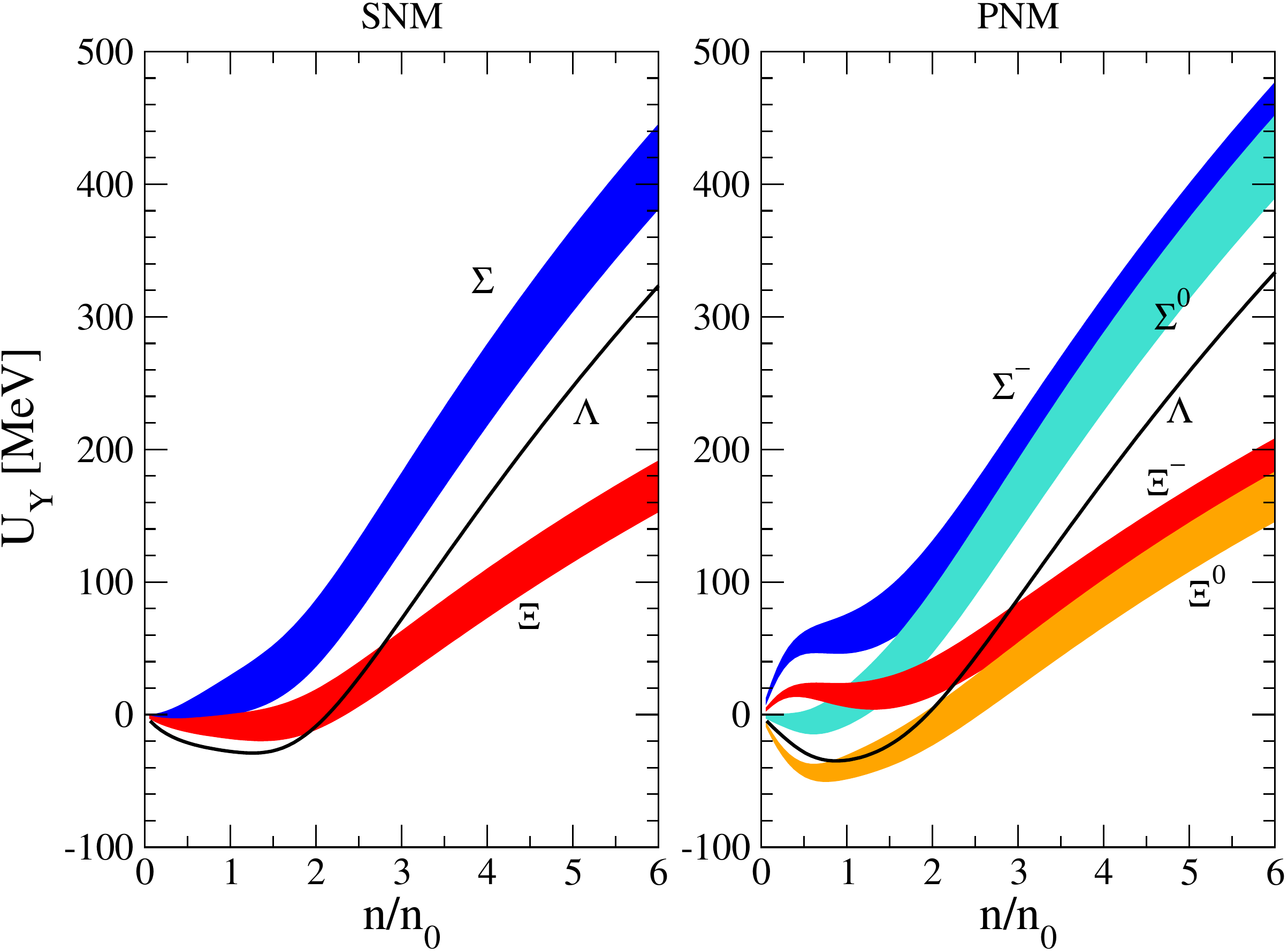}
\caption{Hyperon single-particle potentials of our RMF models, as functions of the nuclear density, in the case of isospin SNM (left panel) and PNM (right panel). The bands result from varying the hyperon-$\sigma$ couplings within the values given in Eq.~(\ref{eq:sigma}) to account for the experimental uncertainties of the hyperon potentials derived from hypernuclear data.}
\label{fig:uhyp}
\end{center}
\end{figure}

The hyperon potentials are shown, as functions of the nuclear density, in Fig.~\ref{fig:uhyp}. The left panel shows the potentials for isospin SNM, while the right panel corresponds to PNM, which is closer to the conditions of beta stable neutron star matter, where differences between the potentials for the different members of the same isospin multiplet can be seen. In order not to overcrowd the figure we have omitted the potentials of the positively charged hyperons, as they do not appear in the beta stable neutron star matter configurations appropriate for the present study. The coloured bands enclose the dispersion of results obtained employing the hyperon-$\sigma$ coupling ranges displayed in Eq.~(\ref{eq:sigma}). The range of couplings has been determined from the uncertainties of the hypernuclear data and, strictly speaking, corresponds to normal nuclear matter density, $n_0$. However, the coupling constants are density independent in our model and we can then obtain a range of values for the potential at any density. Specially important are the potentials around $2n_0$ and beyond, which is the region of densities where hyperons are present in the models explored here \citep{Tolos:2016hhl}. As can be seen from the figure, the range 
of values for the hyperon potentials at $2n_0$ in PNM are the following: $U_{\Lambda}^{(N)}(2n_0)=4$ MeV, $U_{\Sigma^-}^{(N)}(2n_0)=84~{\rm to}~130$ MeV, $U_{\Sigma^0}^{(N)}(2n_0)=47~{\rm to}~93$ MeV, $U_{\Xi^-}^{(N)}(2n_0)=14~{\rm to}~42$ MeV and $U_{\Xi^0}^{(N)}(2n_0)=-22~{\rm to}~5$ MeV. We note that this  range of values will strongly affect the composition of the neutron star, as we will show at the end of this section.

 \begin{figure}[t!]
\begin{center}
\includegraphics[width=0.48\textwidth]{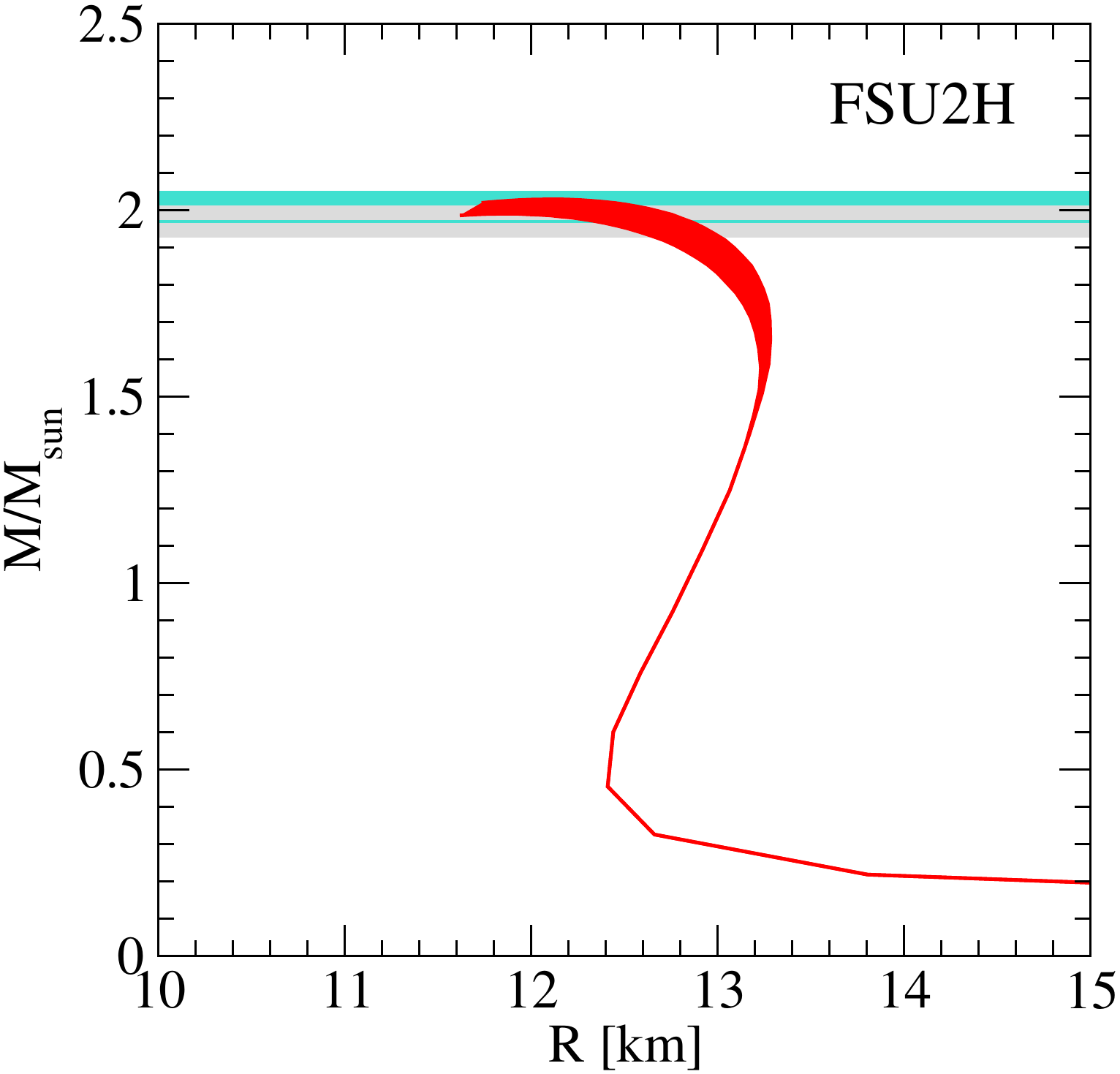}
\caption{Mass versus radius for neutron stars from the FSU2H model. The band results from varying the hyperon-$\sigma$ couplings within the values given in Eq.~(\ref{eq:sigma}) to account for the experimental uncertainties of the hyperon potentials derived from hypernuclear data.}
\label{fig:M-R_hyp_sigma_coup}
\end{center}
\end{figure}

In Fig.~\ref{fig:M-R_hyp_sigma_coup} we show the mass-radius relation for neutron stars obtained with the FSU2H model. The band collects the results obtained varying the hyperon couplings to the $\sigma$ meson within the ranges in Eq.~(\ref{eq:sigma}), which produce maximum masses that differ by at most  $0.1 M_\odot$. This is a small effect, as it is obvious that the hyperon potentials at nuclear densities of around $6n_0$ in the center of $2 M_\odot$ stars (see Table \ref{tab:starprops}) suffer a much larger uncertainty than the one we extrapolated from the normal nuclear densities characteristic of hypernuclear data. Indeed, the uncertainties tied to our lack of knowledge of the hyperon-nucleon and hyperon-hyperon interactions around the hyperon onset density of  $\sim 2n_0$ and beyond have often been exploited to build up RMF models that produce hyperonic neutron stars with maximum masses larger than $2 M_\odot$  \citep{Weissenborn:2011ut,Bednarek:2011gd,vanDalen:2014mqa,Oertel:2014qza,Fortin:2017cvt}. As can be seen in the extensive analyses of various models in \citep{Fortin:2014mya}, the maximum masses turn out to be within a  $0.3 M_\odot$ band. 
It is therefore clear that determining the hyperon interactions at higher densities, as could be done from the analysis of  HIC experiments \citep{Morita:2014kza}, would help constraining the models in the appropriate regimes found in neutron stars.
 
\begin{figure}[ht]
\begin{center}
\includegraphics[width=0.45\textwidth]{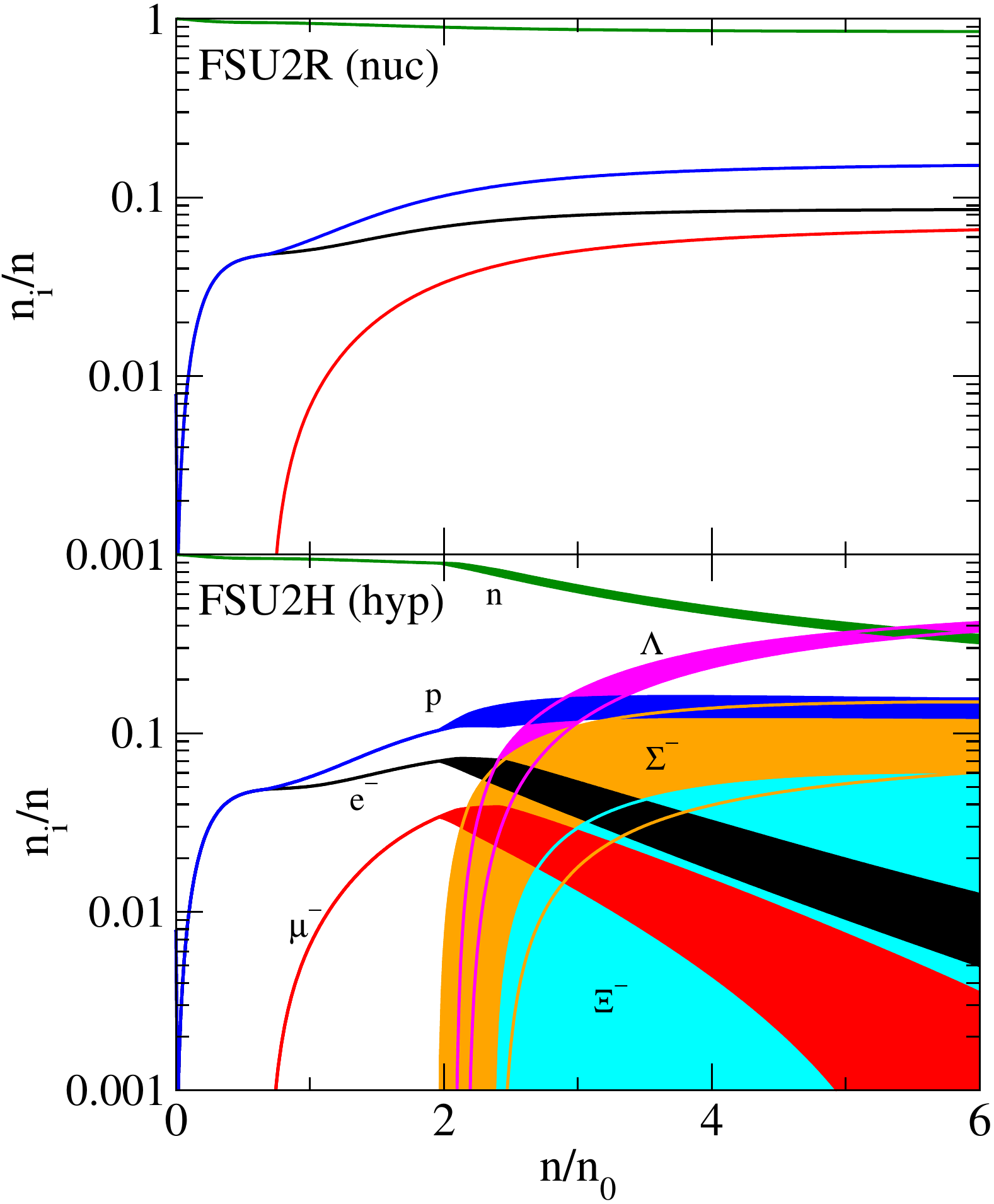}
\caption{Particle fractions as functions of the baryonic density for the nucleonic FSU2R model (upper panel) and the hyperonic FSU2H model (lower panel). The bands in the lower panel result from varying the hyperon-$\sigma$ couplings within the values given in Eq.~(\ref{eq:sigma}) to account for the experimental uncertainties of the hyperon potentials derived from hypernuclear data. The coloured lines guide the eye to help distinguishing each case properly in the regions of overlapping bands.}
\label{fig:fractions}
\end{center}
\end{figure}

Let us finish this section by showing, in Fig.~\ref{fig:fractions}, the particle fractions as functions of the baryonic density, for the FSU2H model (lower panel), where the coloured bands are obtained for the range of hyperon-$\sigma$ couplings employed in this work. For completeness, we also show the particle fractions for the nucleonic FSU2R model in the upper panel, where we can see that the absence of negatively charged hyperons maintains a constant population of electrons and muons, and hence of protons and neutrons, already from slightly above $2n_0$. As for the FSU2H model, we note that all the particle fractions are affected by the hypernuclear data uncertainties, even if these are encoded only in the $\Sigma$ and $\Xi$ couplings to the $\sigma$ meson. Upon inspecting the range of densities where hyperons may be present, we see that, although one can generally conclude that hyperons appear around $2n_0$,  the order of appearance of each species is not determined, owing to the uncertainties derived from hypernuclear data. The first hyperon to appear can be either a $\Lambda$ or a $\Sigma^-$, the latter case only in the less repulsive situation allowed by data, namely when $U_{\Sigma}^{(N)}(n_0)\simeq 0$~MeV. In fact, when the $\Sigma$ feels its most repulsive potential value, it can even appear after the $\Xi^-$ hyperon. This happens when the $\Xi$ potential value is on the most attractive side of the allowed region, namely  $U_{\Sigma}^{(N)}(n_0)=-18$~MeV. However, if one decreases the amount of attraction, as data permits, the $\Xi^-$ onset density is rapidly pushed towards larger values, even beyond the maximum density of $6n_0$ represented in the figure, which stands as a representative central density of hyperonic stars. 
Summarizing, although hyperons are present in the interior of neutron stars modeled by the FSU2H interaction, the lack of precise knowledge on the hyperon-nuclear interactions prevents one from establishing the specific hyperonic composition in the interior of the star.

\section{Summary}
\label{sec:summary}

We have reinvestigated our previous results on the equation of state for the nucleonic and hyperonic inner core of neutron stars  \citep{Tolos:2016hhl}, that 
fullfill the 2$M_{\odot}$ observations \citep{Demorest:2010bx,Antoniadis:2013pzd} and the recent determinations of radii below 13 km region  \citep{Guillot:2013wu,Lattimer:2013hma,Heinke:2014xaa,Guillot:2014lla,Ozel:2015fia,Lattimer:2015nhk}, as well as the saturation properties of nuclear matter and finite nuclei \citep{Tsang:2012se,Chen:2014sca} and the constraints extracted from HICs \citep{Danielewicz:2002pu, Fuchs:2000kp,Lynch:2009vc}. The two models formulated in  \citep{Tolos:2016hhl}, the FSU2R (with nucleons) and FSU2H (with nucleons and hyperons) models, have been updated by improving the behavior of PNM at subsaturation densities. Above saturation density, the updated models and those of \citep{Tolos:2016hhl} are very similar. 

Using these updated interactions, we have obtained values for the PNM pressure at saturation density of 2.44 MeV fm$^{-3}$ in FSU2R and of 2.30 MeV fm$^{-3}$ in FSU2H, that are consistent with the estimates from chiral forces \citep{Fortin:2014mya, Hagen:2015yea}. The symmetry energy and its slope at saturation become $E_{\rm sym}(n_0)= 30.7$~MeV and $L= 46.9$~MeV in FSU2R and $E_{\rm sym}(n_0)= 
30.5$~MeV and $L= 44.5$~MeV in FSU2H, thus being in good agreement with several recent estimates based on terrestrial experiments, different astrophysical observations, and theoretical calculations \citep{Li:2013ola,Lattimer:2012xj,Roca-Maza:2015eza,Hagen:2015yea,Oertel:2016bki,Birkhan:2016qkr}. Furthermore, 
the reviewed FSU2R and FSU2H models predict a neutron skin thickness of 0.15 fm in $^{208}$Pb and of 0.166 fm in $^{48}$Ca, which turn out to be compatible with previous experimental and theoretical determinations \citep{Roca-Maza:2015eza,Tarbert:2013jze,Abrahamya12,Horowitz:2012tj,Hagen:2015yea,Birkhan:2016qkr} 

With regards to the mass and radius of neutron stars, radii below 13 km can be achieved because of the softening of the symmetry energy around saturation density whereas, at the same time, $2 M_\odot$ stars can be obtained as the pressure of the high-density EoS is hard enough. These results  are not drastically changed when using the updated FSU2R and FSU2H interactions as compared to the previous versions in Ref.~\citep{Tolos:2016hhl}, because of the similar EoSs produced above saturation density. The numerical tabulations of the EoS and of the M-R relation from the models FSU2R ($npe\mu$ matter) and FSU2H ($npY\!e\mu$ matter) as a function of the number density $n/n_0$ are shown in Tables \ref{tab:eosdata1} and \ref{tab:eosdata2} for completeness.

However, the mass and composition of neutron stars might be strongly affected  due to the uncertainties of the hyperon-nucleon couplings. The values of the hyperon couplings are determined from SU(3) flavor symmetry and from the available experimental information on hypernuclei, in particular by fitting to the optical potential of hyperons extracted from the data. The coupling of each hyperon to the $\sigma$ meson field is left as a free parameter to be adjusted to reproduce the hyperon potential in SNM within the experimental uncertainties. As a result, we have found that the onset of appearance of the different hyperons strongly depends on the hyperon-nuclear uncertainties, whereas the maximum masses differ by at most 0.1 $M_{\odot}$, thus being less sensitive to the changes on the hyperon-nucleon couplings. This latter conclusion has to be taken with care, though, since the hyperon potentials at densities in the center of $2 M_\odot$ stars suffer much larger uncertainties than the ones we have extrapolated from  hypernuclear data at saturation. Hence,  a greater dispersion of values for the maximum mass might be expected. The progress in the characterization of hyperon-nucleon interactions in dense matter derived from chiral effective forces \citep{Haidenbauer:2016vfq}, on the theoretical front, and from studies of HICs \citep{Morita:2014kza}, on the experimental front,  
should contribute greatly to narrow down these uncertainties.

\begin{acknowledgements}
L.T. acknowledges support from the Ram\'on y Cajal research programme,
FPA2013-43425-P  and FPA2016-81114-P Grants from Ministerio de Economia y Competitividad (MINECO) and NewCompstar COST Action MP1304.
M.C. and A.R. acknowledge support from Grant No. FIS2014-54672-P from MINECO, Grant No. 2014SGR-401 from Generalitat de Catalunya, and the project MDM-2014-0369 of ICCUB (Unidad de Excelencia Mar\'{\i}a de Maeztu) from MINECO.
L.T. and A.R. acknowledge support from the Spanish Excellence Network on Hadronic Physics FIS2014-57026-REDT from MINECO.
\end{acknowledgements}

\bibliographystyle{pasa-mnras}
\bibliography{biblio_rev}

\newpage

\appendix

\section*{Appendix} 
 
 For ease of use, in this appendix we provide in tabular form the results for the EoS and the M-R relation 
calculated with the FSU2R and FSU2H models discussed in the text.

\begin{table*}
\caption{Numerical data of the EoS for the core of neutron stars and of the M-R relation from the models FSU2R ($npe\mu$ matter) and FSU2H ($npY\!e\mu$ matter), 
as a function of the number density $n/n_0$ (with $n_0=0.1505$~fm$^{-3}$, cf. Table \ref{t-props}). The pressure $P$ and mass-energy density $\varepsilon$ are in MeV fm$^{-3}$, while the neutron star radius $R$ 
and mass $M$ are in km and $M_\odot$ units, respectively.}
\centering
\begin{tabular*}{\textwidth}{\x l cccc cccc}
\hline\hline
 & \multicolumn{4}{c}{FSU2R model (nucleonic)} & \multicolumn{4}{c}{FSU2H model (hyperonic)} \\
   \cline{2-5}                            \cline{6-9}
$n/n_0$ & $P$ & $\varepsilon$ & $R$ & $M$  & $P$ & $\varepsilon$ & $R$ & $M$ \\
\hline
       0.9    &    1.3737    &    128.98    &    28.701    &     0.10    &    1.2675    &    128.98    &    36.393    &     0.09\\
       1.0    &    2.1255    &    143.50    &    17.229    &     0.15    &    2.0182    &    143.49    &    18.143    &     0.14    \\
       1.1    &    3.0982    &    158.11    &    14.256    &     0.21    &    3.0319    &    158.08    &    14.317    &     0.20    \\
       1.2    &    4.3089    &    172.82    &    13.129    &     0.27    &    4.3426    &    172.79    &    13.015    &     0.27    \\
       1.3    &    5.7741    &    187.63    &    12.660    &     0.35    &    5.9878    &    187.61    &    12.541    &     0.36    \\
       1.4    &    7.5095    &    202.58    &    12.482    &     0.43    &    8.0082    &    202.58    &    12.400    &     0.46    \\
       1.5    &    9.5289    &    217.65    &    12.428    &     0.52    &    10.445    &    217.70    &    12.416    &     0.57    \\
       1.6    &    11.844    &    232.87    &    12.440    &     0.61    &    13.340    &    233.00    &    12.509    &     0.68    \\
       1.7    &    14.462    &    248.24    &    12.498    &     0.70    &    16.725    &    248.50    &    12.638    &     0.81    \\
       1.8    &    17.389    &    263.78    &    12.553    &     0.79    &    20.626    &    264.21    &    12.764    &     0.93    \\
       1.9    &    20.626    &    279.49    &    12.618    &     0.88    &    25.058    &    280.16    &    12.885    &      1.06    \\
       2.0    &    24.171    &    295.37    &    12.680    &     0.97    &    30.022    &    296.35    &    13.004    &      1.18    \\
       2.1    &    28.019    &    311.44    &    12.739    &      1.06    &    35.510    &    312.80    &    13.099    &      1.30    \\
       2.2    &    32.162    &    327.70    &    12.774    &      1.14    &    41.504    &    329.52    &    13.175    &      1.41    \\
       2.3    &    36.592    &    344.16    &    12.813    &      1.22    &    46.755    &    346.51    &    13.233    &      1.49    \\
       2.4    &    41.297    &    360.81    &    12.829    &      1.29    &    51.366    &    363.71    &    13.255    &      1.56    \\
       2.5    &    46.266    &    377.67    &    12.844    &      1.36    &    55.951    &    381.10    &    13.274    &      1.61    \\
       2.6    &    51.487    &    394.73    &    12.851    &      1.42    &    60.285    &    398.67    &    13.282    &      1.66    \\
       2.7    &    56.947    &    411.99    &    12.854    &      1.48    &    64.452    &    416.40    &    13.284    &      1.69    \\
       2.8    &    62.634    &    429.47    &    12.843    &      1.53    &    68.763    &    434.29    &    13.272    &      1.73    \\
       2.9    &    68.536    &    447.14    &    12.830    &      1.58    &    73.242    &    452.33    &    13.269    &      1.76    \\
       3.0    &    74.642    &    465.03    &    12.816    &      1.63    &    77.898    &    470.53    &    13.239    &      1.78    \\
       3.1    &    80.941    &    483.12    &    12.792    &      1.67    &    82.732    &    488.89    &    13.224    &      1.81    \\
       3.2    &    87.421    &    501.42    &    12.763    &      1.71    &    87.745    &    507.41    &    13.194    &      1.83    \\
       3.3    &    94.075    &    519.92    &    12.734    &      1.74    &    92.934    &    526.09    &    13.161    &      1.85    \\
       3.4    &    100.89    &    538.63    &    12.702    &      1.77    &    98.299    &    544.93    &    13.134    &      1.87    \\
       3.5    &    107.87    &    557.54    &    12.669    &      1.80    &    103.84    &    563.93    &    13.093    &      1.89    \\
       3.6    &    114.99    &    576.66    &    12.633    &      1.83    &    109.55    &    583.09    &    13.058    &      1.90    \\
       3.7    &    122.25    &    595.97    &    12.603    &      1.85    &    115.42    &    602.41    &    13.016    &      1.92    \\
       3.8    &    129.64    &    615.48    &    12.566    &      1.87    &    121.46    &    621.89    &    12.982    &      1.93    \\
       3.9    &    137.16    &    635.18    &    12.521    &      1.89    &    127.66    &    641.53    &    12.935    &      1.94    \\
\hline\hline
\end{tabular*}
\label{tab:eosdata1}
\end{table*}

\begin{table*}[t!]
\caption{Continuation of Table \ref{tab:eosdata1}.}
\centering
\begin{tabular*}{\textwidth}{\x l cccc cccc}
\hline\hline
 & \multicolumn{4}{c}{FSU2R model (nucleonic)} & \multicolumn{4}{c}{FSU2H model (hyperonic)} \\
   \cline{2-5}                            \cline{6-9}
$n/n_0$ & $P$ & $\varepsilon$ & $R$ & $M$  & $P$ & $\varepsilon$ & $R$ & $M$ \\
\hline
      4.0    &    144.80    &    655.08    &    12.479    &      1.91    &    134.02    &    661.33    &    12.886    &      1.95    \\
       4.1    &    152.56    &    675.18    &    12.442    &      1.93    &    140.53    &    681.30    &    12.842    &      1.96    \\
       4.2    &    160.44    &    695.46    &    12.400    &      1.94    &    147.19    &    701.42    &    12.804    &      1.97    \\
       4.3    &    168.41    &    715.93    &    12.359    &      1.96    &    154.00    &    721.71    &    12.756    &      1.98    \\
       4.4    &    176.49    &    736.59    &    12.323    &      1.97    &    160.96    &    742.15    &    12.703    &      1.99    \\
       4.5    &    184.67    &    757.44    &    12.280    &      1.98    &    168.06    &    762.76    &    12.662    &      1.99    \\
       4.6    &    192.95    &    778.46    &    12.235    &      1.99    &    175.30    &    783.52    &    12.613    &      2.00    \\
       4.7    &    201.32    &    799.67    &    12.196    &      2.00    &    182.67    &    804.45    &    12.571    &      2.00    \\
       4.8    &    209.77    &    821.06    &    12.155    &      2.01    &    190.18    &    825.53    &    12.523    &      2.01    \\
       4.9    &    218.31    &    842.62    &    12.124    &      2.01    &    197.81    &    846.77    &    12.476    &      2.01    \\
       5.0    &    226.94    &    864.36    &    12.074    &      2.02    &    205.58    &    868.16    &    12.427    &      2.01    \\
       5.1    &    235.64    &    886.27    &    12.035    &      2.02    &    213.47    &    889.72    &    12.383    &      2.02    \\
       5.2    &    244.43    &    908.36    &    11.996    &      2.03    &    221.48    &    911.42    &    12.339    &      2.02    \\
       5.3    &    253.28    &    930.61    &    11.963    &      2.03    &    229.62    &    933.29    &    12.293    &      2.02    \\
       5.4    &    262.22    &    953.03    &    11.921    &      2.04    &    237.87    &    955.31    &    12.252    &      2.02    \\
       5.5    &    271.22    &    975.62    &    11.886    &      2.04    &    246.24    &    977.48    &    12.208    &      2.02    \\
       5.6    &    280.30    &    998.37    &    11.852    &      2.04    &    254.72    &    999.81    &    12.166    &      2.02    \\
       5.7    &    289.44    &    1021.3    &    11.808    &      2.04    &    263.31    &    1022.3    &    12.120    &      2.02    \\
       5.8    &    298.65    &    1044.4    &    11.775    &      2.04    &    272.01    &    1044.9    &    12.078    &      2.02    \\
       5.9    &    307.93    &    1067.6    &    11.743    &      2.05    &    280.82    &    1067.7    &    12.038    &      2.02    \\
       6.0    &    317.26    &    1091.0    &    11.704    &      2.05    &    289.74    &    1090.6    &    11.997    &      2.02    \\
       6.1    &    326.67    &    1114.5    &    11.670    &      2.05    &    298.76    &    1113.7    &    11.952    &      2.02    \\
       6.2    &    336.13    &    1138.2    &    11.639    &      2.05    &    307.88    &    1136.9    &    11.920    &      2.02    \\
       6.3    &    345.65    &    1162.1    &    11.604    &      2.05    &    317.10    &    1160.3    &    11.872    &      2.02    \\
       6.4    &    355.23    &    1186.1    &    11.573    &      2.05    &    326.42    &    1183.8    &    11.833    &      2.02    \\
       6.5    &    364.87    &    1210.3    &    11.538    &      2.05    &    335.83    &    1207.5    &    11.793    &      2.02    \\
\hline\hline
\end{tabular*}
\label{tab:eosdata2}
\end{table*}

\end{document}